\documentclass[9pt,preprint2,natbib209]{aastex}
\shorttitle{Origin of Dust around V1309 Sco} \shortauthors{Zhu
et al.}

\begin{document}


\title{Origin of Dust around V1309 Sco}
\author{Chunhua Zhu\altaffilmark{1}, Guoliang L\"{u}\altaffilmark{1^\dagger}, Zhaojun Wang\altaffilmark{1}}
\email{$^\dagger$guolianglv@gmail.com}
\altaffiltext{1}{School of Physical Science and Technology, Xinjiang
University, Urumqi, 830046, China.}





\begin{abstract}
The origin of dust grains in the interstellar medium is still open problem. \cite{Nicholls2013} found
the presence of a significant amount of dust around V1309 Sco which maybe originate from the merger of a contact binary.
We investigate the origin of dust around V1309 Sco,
and suggest that these dust grains are efficiently produced in the binary-merger ejecta. By means of \emph{AGBDUST} code, we estimate
that $\sim 5.2\times10^{-4} M_\odot$ of dust grains are produced, and their radii are $\sim 10^{-5}$ cm. These dust grains
 mainly are composed of silicate and iron grains. Because
the mass of the binary-merger ejecta is very small, the contribution of dust produced by binary-merger ejecta to the overall dust
production in the  interstellar medium is negligible.
However, it is the most important that the discovery of
a significant amount of dust around V1309 Sco offers a direct support for the idea---common-envelope ejecta provides an ideal environment for dust
formation and growth. Therefore, we confirm that common-envelope ejecta can be important source of cosmic dust.
\end{abstract}


\section{Introduction}
One of the important constituents of the interstellar medium
(ISM) is dust which plays a crucial role in the astrophysics of the ISM, from
the thermodynamics and chemistry of the gas to the dynamics of star
formation. The origin of cosmic dust is still an open problem.
According to popular point of view,
dust mainly originates from the stellar wind of asymptotic giant branch
(AGB) stars \citep{Gail2009}, or supernova (SN) ejecta \citep{Dunne2003}.  However,
\cite{Draine2009} estimated that the dust originating from AGB stars and supernova could be only
10\% of interstellar dust if their lifetimes are considered. Similarly, in the Large and Small Magellanic Clouds,
\cite{Matsuura2009,Matsuura2013} and \cite{Boyer2012} found that the accumulated dust mass from AGB stars and possibly SNe
is significant less than the dust mass in the ISM, and they suggested that dust must
be formed by another unknown mechanism or grow in the ISM which was put forward by \cite{Draine1979}.
Unfortunately, the destruction and growth processes of dust in the ISM are poorly understood \citep[e.g.,][]{Dwek2007}.

According to
the classical theory of nucleation \citep{Feder1966}, the saturation
pressure in most cases falls more rapidly than the pressure of the
vapour, and the vapour becomes supersaturated when saturated vapour
expands adiabatically. The formation of dust grains from the gas phase
can occur from vapour in a supersaturated state. Therefore, some authors suggested
that the ejecta during common-envelope (CE) evolution in
close binary systems can provide a good environment for dust formation and growth \citep[e.g.,][]{Lu2013,Ivanova2013}.
Compared to AGB stars, \cite{Lu2013} found that the dust quantities produced by CE ejecta may be
significant or even dominated. Thus,  it can be seen that the majority of dust in ISM may originate from CE ejecta.
However, there is no direct evidence on observations to support their
model.

Very recently, \cite{Nicholls2013} showed that V1309 Sco had become dominated by mid-IR emission since eruption,
which indicated the presence of a significant amount of dust in the circumstellar environment.
V1309 Sco erupted in September 2008 \citep{Nakano2008}. Subsequently, its evolution marked
it as a new member of the 'red novae' \citep{Mason2010}. \cite{Soker2003} suggested that red novae
were produced by binary merger. \cite{Tylenda2011} analysed the data of
V1309 Sco photometrically observed by the OGLE project since August 2001, and showed that it
indeed originated from the merger of contact binary. Up to now, V1309 Sco is the first documented case of
a binary merger. Furthermore, having considered that the absence of
detectable mid-IR emission before the outburst,
\cite{Nicholls2013} suggested
that the dust around V1309 Sco was produced in the eruptive merger event.

In general, before a binary merger, the binary undergoes a CE phase which
forms as a result of dynamical timescale mass
exchange in close binaries and plays an essential role in their
evolution \citep[see, e.g.,][]{Paczynski1976,Iben1993}. In most cases,
CE evolution involves a giant star (donor) transferring matter to a main sequence star or
a degenerate dwarf (gainer) on a dynamical timescale. The giant envelope
overfills the Roche lobes of both stars and engulfs the giant core and
its companion, and forms a CE. During the CE phase, owing to its expansion the CE rotates more slowly
than the orbit velocity of the giant core and the donor, which results in friction occurring.
Then, orbital energy is transferred to
the CE via dynamical friction between the orbiting components and the
non-corotating CE. If the orbital energy is enough large, whole CE can be ejected on a
dynamical timescale, the binary does not merge and a close binary is left. In this work,
the whole envelope ejected is called as CE ejecta. The CE ejecta rapidly expands, and its temperature rapidly decreases.
\cite{Lu2013} investigated that dust forms and grows in it.

If the orbital energy is too small, the gainer and the donor merge into a single
star, and part of CE may be ejected. In order to distinguish from CE ejecta,
we call the matter ejecta in this case as binary-merger ejecta.
Usually, compared with whole CE mass, the mass of a binary-merger ejecta is small.
Using the 3D SPH code \emph{StarCrash}, \cite{Ivanova2013} performed several numerical simulations
to estimate the mass ejected during the merging process of the progenitor binary of V1309 Sco. In their
simulations the merger ejects a small fraction of the giant envelope, and the ejected mass varies from 0.03 to 0.08
$M_\odot$. \cite{Ivanova2013} suggested that the binary-merger ejecta rapidly cooled down, and potentially formed dust around V1309 Sco.
In present work, we try to simulate
the dust formation and growth in the binary-merger ejecta around V1309 Sco.
If the merger event in V1309 Sco indeed produced a significant amount of dust,
it means that dust can efficiently form and grow in the binary merger ejecta. Similarly, dust can
do so in the CE ejecta.
Therefore, observed dust around the V1309 Sco
offers a support for the new origin of cosmic dust, which was proposed by \cite{Lu2013}.

In this paper, we explain the origin of dust around V1309 Sco and show the quantity and properties of these dust grains.
In \S 2 the model of the ejecta during binary merger is described.  Results and
discussions are given in \S  3.  \S  4
gives conclusions.

\section{Dust Model for V1309 Sco}
In order to simulate the formation and growth of dust around V1309 Sco, we must
know its progenitor which was composed of a gainer and a donor.
\cite{Tylenda2011} estimated the luminosity and effective temperature of the progenitor
of V1309 Sco, and suggested that the donor was at the beginning of the red giant branch.
According to the observational constraints,
\cite{Stepien2011} computed the evolution of the progenitor binary of V1309 Sco from zero-age main
sequence to binary merger. He showed that all possible progenitor systems had undergone
the second Roche lobe overflow before merger.  At this time, the donor was a giant star and it's mass was $\sim$1.5 $M_\odot$;
the gainer's mass was $\sim$ 0.15 $M_\odot$ and binary orbital period was $\sim$ 1.4 days.

Based on the popular model of binary evolution, if the donor is overflowing its Roche lobe, the
progenitor will undergo a CE evolution and the donor and the gainer finally merge into V1309 Sco.
During the above progress, a portion of the donor's envelope is ejected. Dust formation and growth
mainly depend on the mass, temperature and mass density of the ejecta.

\subsection{Mass of Ejecta}
At the beginning of the giant branch, the donor in the progenitor binary of V1309 Sco had
an envelope with a mass of $\sim 1.2 M_\odot$. As the last section described, \cite{Ivanova2013}
estimated that the ejected mass during the binary merging process varies from 0.03 to 0.08
$M_\odot$.

Considering that a significant amount dust around V1309 Sco was found, we adopt an upper limit of
their estimate, that is, the ejected mass equals 0.08 $M_\odot$.
Because of the ejecting process being on a dynamical timescale, we assume that the most outer matter of
donor's envelope is ejected, which is showed by the right region of dotted line in Fig. \ref{fig:tede}.

\subsection{Mass Density and Temperature of Ejecta}
\begin{figure}
\includegraphics[totalheight=3.3in,width=3.0in,angle=-90]{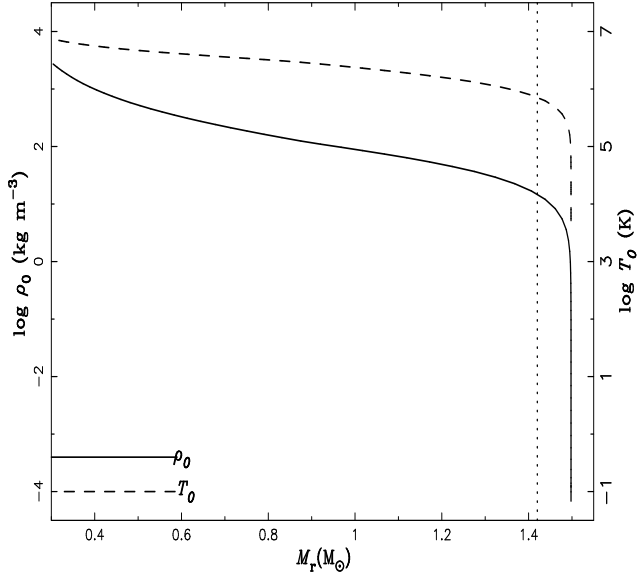}
\caption{The mass density and the temperature of the donor' envelope in the progenitor of
V1309 Sco at the beginning of binary merger. The envelope on the right region of dotted line
is ejected during merging process. }\label{fig:tede}
\end{figure}

The ejecta arising from binary merger immediately expands, and its mass density and temperature begin to decrease.
It is possible that dust formation and growth occur in the ejecta.
To our knowledge, there is no any comprehensive theoretical model or observational datum to
describe the mass density and the temperature of ejecta after binary merger.
Considering that binary merger and ejecting matter occur within dynamical timescale, we assume that
the ejecta mainly originates from a part of donor's envelope, and its initial mass density
and temperature are showed in Fig. \ref{fig:tede}.

On investigating dust formation
and growth in CE ejecta, \cite{Lu2013} assumed a toy model to simulate the
evolution of mass density and temperature  of CE ejecta.
We think that the evolution of mass density and temperature of binary-merger ejecta is similar to
that of CE ejecta, and we adopt the descriptions in \cite{Lu2013}.

After the matter is ejected, its mass density, $\rho$, is
represented by $\rho={\dot{M}_{\rm ej}}/({4\pi R^2 V})$, where
$\dot{M}_{\rm ej}$ and $R$ are the mass-ejection rate and the radial
distance of the ejected matter, respectively, and $V$ is the velocity
of the ejected matter. We assume V approximately equals to the escape
velocity. Because the matter is ejected on a dynamical timescale, we
assume that $\dot{M}_{\rm
  ej}=\rho_04\pi R_0^2 V_0$, where $R_0$ is the initial radius of the ejected material,
  $V_0$ is the the escape velocity of the ejected matter at $R_0$ and $\rho_0$ is the
initial mass density.
Following  \cite{Lu2013}, the mass density of binary-merger ejecta at $R$ can be approximated by
\begin{equation}
\rho=(\frac{R_0}{R})^{3/2}\rho_0,\label{eq:rho}.
\end{equation}

\cite{Lu2013} considered that the expansion of CE ejecta underwent two different zones in which
the evolution of temperature is different.
\\ (i)In the first zone, some hydrogen atoms are ionized.
With the temperature decreases, ionized hydrogens gradually turn
into hydrogen atoms via releasing certain energy.
The released energy can partly be absorbed by CE ejecta, partly be used to drive
the CE \citep{Han1995b,Han1995a} or directly is lost from CE ejecta. For simplicity, \cite{Lu2013}
introduced a parameter $\gamma$ to describe the temperature
evolution:
\begin{equation}
T=(\frac{R_0}{R})^{\gamma}T_0, \label{eq:1zone}
\end{equation}
where $T_0$ is the initial temperature. The ionization of hydrogen atom can be given
by Saha ionization equation.
\\ (ii)In the second zone, the majority of ionized
hydrogens have turned into hydrogen atoms, the gas in CE ejecta is similar
with that in the surface of red giant. \cite{Lu2013}
selected the temperature's evolution given by \cite{Lucy1971,Lucy1976}
via
\begin{equation}
T^4=\frac{1}{2}T^4_{\rm cr}(1-\sqrt{1-\frac{R^2_{\rm
cr}}{R^2}}+\frac{3}{2}\tau_{\rm L}) \label{eq:nate}
\end{equation}
where $\tau_{\rm L}$ is defined by
\begin{equation}
\frac{{\rm d}\tau_{\rm L}}{{\rm d}r}=-\rho \kappa_{\rm
H}\frac{R^2_{\rm cr}}{R^2}.
\end{equation}
Where, $\kappa_{\rm H}$ is the flux averaged mass extinction
coefficient, and is calculated by a simple
superposition of the extinction of the different dust species and
the gas (See details in \cite{Gail1999}).  Here, $R_{\rm cr}$ is the boundary
of two zone where the ion degree of hydrogen atoms equals 1\%, and $T_{\rm cr}$ is
the temperature of CE ejecta at $R_{\rm cr}$.

Because of the lack of observational data for temperature evolution of CE ejecta,
it is difficult to determine parameter $\gamma$. \cite{Elvis2002} investigated
the dust formation in an outflowing wind from quasars. They assumed that the
outflowing is an ideal gas obeying polytropic adiabatic scaling and further
that the gas is monoatomic with index $\gamma_{\rm ad}=5/3$. In their model,
the wind cooling law is given by $T/T_0=(\frac{R_0}{R})^{0.5}$.
Based on the results of model calculations by \cite{Fransson1989},
\cite{Kozasa1989} adopted an adiabatic index $\gamma_{\rm ad}=1.25$
for the early stage of SN explosions. For an adiabatically expanding perfect
gas, $\rho T^{\frac{1}{1-\gamma_{\rm ad}}}={\rm constant}$.
If the temperature evolution of the CE ejecta in
the inner zone is similar to that in the early stage of a SN explosion.
This implies $\gamma\sim0.4$ in Eq. (\ref{eq:1zone}).
In order to check parameter $\gamma$ effects on dust formation, \cite{Lu2013}
carried out different simulations in which $\gamma=0.2$, 0.3 and 0.4, respectively.
They found that CE ejecta could efficiently produce dust in the simulation
with $\gamma = 0.4$.  We also carried out a test: If
 $\gamma= 0.5$, most of Si, Fe and Mg elements condense into the
dust grains of silicate and iron. Considering that \cite{Nicholls2013} found  a significant
amount of dust in the circumstellar environment around V1309 Sco,
we take $\gamma=0.4$ in this work.

\subsection{Dust Formation and Growth}
Using \emph{AGBDUST} code, \cite{Gail1999} had
investigated the condensation and growth of dust grains in the
stellar wind from AGB star \citep[Also see][]{Ferrarotti2001,Ferrarotti2002,Ferrarotti2003,Ferrarotti2005}.
As the last section mentioned, the binary-merger ejecta in the second zone is similar to
the stellar wind from AGB star. We use \emph{AGBDUST} code to simulate the
dust formation and growth in the binary-merger ejecta. In the \emph{AGBDUST} code,
for a gas given temperature and mass density (The temperature and the mass density of binary-merger ejecta
are determined by Eqs.(\ref{eq:rho}), (\ref{eq:1zone}) and (\ref{eq:nate}).), the  chemical
abundances affect the dust species.

For a single star, three dredge-up processes and
hot bottom burning in a star with initial mass higher than
$4M_\odot$ may change the chemical abundances of the stellar
envelope\citep{Iben1983}. In binary systems, mass transfer can change the chemical abundances
of stellar envelope. \cite{Stepien2011} showed that the progenitor of V1309 Sco had undergone two Roche lobe overflows
and the giant before binary merger had accreted a large amount of matter ($\sim 0.6$---1.0 $M_\odot$) from its companion at
the first Roche lobe overflow. According to the models of \cite{Stepien2011}, the giant and its companion in the progenitor
of V1309 Sco had only undergone the first dredge-up. \cite{Iben1983} showed that the
effects of the first dredge-up are a reduction of $^{12}$C by approximately 30\% and no change in the $^{16}$O abundance at the
stellar surface. Other key elements (Fe, Si, Mg and S)
in stellar envelope for dust formation do not change.
 Therefore, if we assume that the initial abundances are similar with those on the surface of the Sun,
 the abundance ratio of carbon element to
oxygen element ($C/O$) in the giant envelope is 0.4 which also is the value of $C/O$ in the binary-merger
ejecta. According to
\cite{Gail1999}, the most abundant dust species formed in the
circumstellar matter ($C/O<1$) are olivine- and pyroxene-type
silicate grains, quartz and iron grains. Before these dust grains start to condense, there
must be some kind of seed nuclei. However, the formation of the seed nuclei are also very difficult problem \citep[e.g.,][]{Gail1999}.
The \emph{AGBDUST} code assumes that the seed nuclei have existed, and their radii are 1 nm \citep{Gail1999,Ferrarotti2006}.
When the temperature of binary-merger ejecta cools down to a limit value, the different elements of dust dust species condensate on the
surfaces of these seed nuclei. The details can be seen in \cite{Gail1999} and \cite{Ferrarotti2006}.

In \emph{AGBDUST} code, besides the temperature, velocity, mass density and chemical abundances, the stellar luminosity and mass can
affect dust formation and growth. However, their effects are weak because
the region of dust formation is far away from the
binary system\citep{Lu2013}. Therefore, we ignore them.
In addition, other input parameters
(Such as outer radius of the zone for dust formation and the number of radial gripdpoints in \emph{AGBDUST} code) which are not specifically
mentioned are taken to have the default values as
in \cite{Ferrarotti2006}.

\section{Results and Discussions}
\begin{figure}
\includegraphics[totalheight=3.3in,width=3.0in,angle=-90]{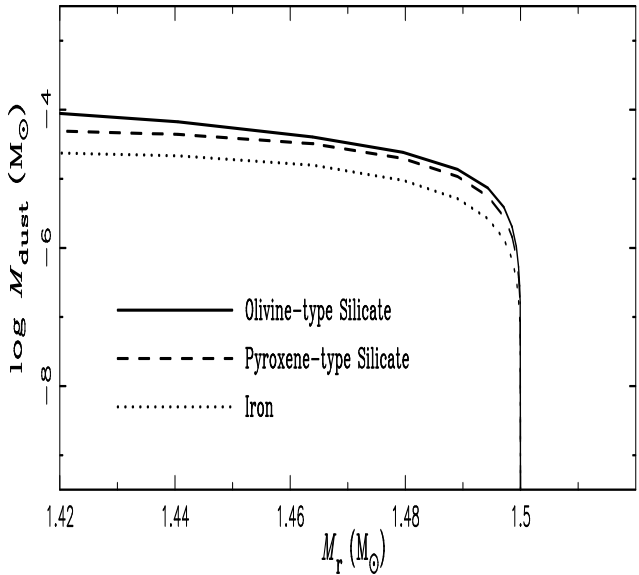}
\caption{The quantities of different dust species along the  mass coordinate of the donor.
  The key to the line-styles representing different dust species is given
            in the bottom left corner. }\label{fig:evequal}
\end{figure}

We simulate the dust formation and growth in the binary-merger ejecta from the progenitor of V1309 Sco.
Their quantities produced along the donor's mass coordinate are showed in Fig. \ref{fig:evequal}.
According to our simulations,  $\sim 5.2\times10^{-4} M_\odot$ of dust grains are produced.
Due to the small sticking efficiency for quartz \citep{Gail1999}, its
quantity in dust grains is negligible. The dust around V1309 Sco
consists of  olivine- and pyroxene-type silicate grains and iron grains.
The proportion of olivine-type silicate grains in whole dust grains is $\sim 50\%$ and
the proportion of pyroxene-type silicate grains is $\sim 30\%$.

Similar to CE ejecta, the mass density ($\sim 3.3\times10^{-11} {\rm g\ cm^{-3}}$) of the binary-merger ejecta in the dust forming zone is
much higher than that ($\sim 5\times10^{-14} {\rm g\ cm^{-3}}$) in an AGB wind\citep{Lu2013}. This condition is very favorable for dust
formation and growth. About 90\% of the Si elements in the binary-merger ejecta
condensate into silicate grains, and about 50\% of the Fe elements condensate into iron grains.

\cite{Nicholls2013} did not estimate total dust mass around V1309 Sco because the
observational data can not allow to estimate the radial extent of the dust.
However, they found two particular results:\\ (i) The temperature of the dust are warm
($\sim 800\pm25$K), which indicates that dust grains form recently;\\ (ii) The best fit to
the absorption features is amorphous pyroxene
grains whose sizes are $\sim$ 3$\times10^{-4}$cm and shape distributions are hollow sphere.

\begin{figure}
\includegraphics[totalheight=3.3in,width=3.0in,angle=-90]{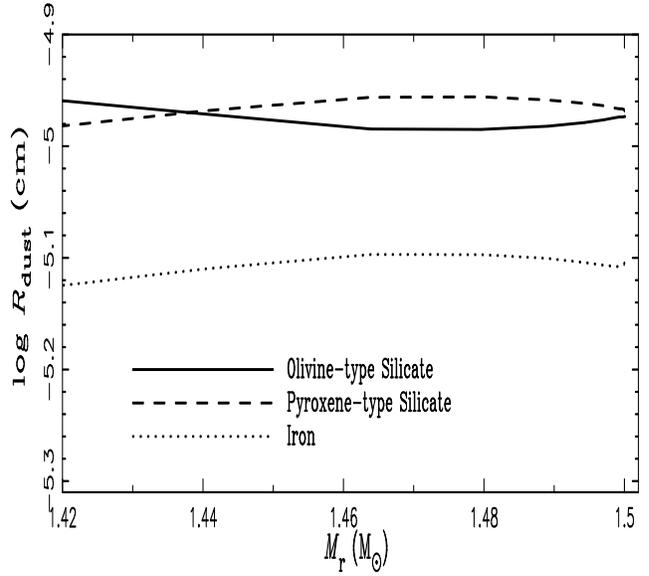}
\caption{The radii of different dust grains along the  mass coordinate of the donor.
     The key to the line-styles representing different dust species is given
            in the bottom left corner.}
\label{fig:everadius}
\end{figure}

In our simulations, the temperature of the region in which dust grains efficiently grow is
$\sim$ 1100---900 K. This is consistent with the observations. However, as Fig. \ref{fig:everadius} shows,
the radii of different dust grains in our simulations are $\sim$0.7---2.8$\times10^{-5}$ cm, which is much smaller than
the observational value.
We consider that reasons are the followings:
\\(i)The \emph{AGBDUST} code assumes that all olivine-type silicate, pyroxene-type silicate and iron grains have
the shape of solid sphere. Given a certain mass, a dust grain with a shape of hollow sphere has a radius larger than
that with a shape of solid sphere.
\\(ii)In \emph{AGBDUST}
code, given the gas's mass density, temperature and chemical compositions, the sizes of dust grains are
affected by an artificial parameter $n_{\rm d}$ which is the number
density of seed nuclei of the different dust species. \cite{Ferrarotti2003} suggested that a lower density
number of seed nuclei seems to give too small dust quantities for the silicates and probably too big silicate
dust particle radii which are bigger than the upper cutoff of the
size distribution derived by \cite{Jura1996}. In order to better agreement with the results in \cite{Jura1996},
\cite{Ferrarotti2003} assumed that $n_{\rm d}=3\times10^{-13}N_{\rm H}$, where $N_{\rm H}$ is the number of hydrogen nuclei.
Correspondingly, the radii of dust grains calculated by \emph{AGBDUST} are $\sim 10^{-5}$cm.
We had carried out different simulations in which  $n_{\rm d}$ is changed from $3\times10^{-13}N_{\rm H}$ to $10^{-15}N_{\rm H}$.
The radii of different dust grains increase up to $5\times10^{-5}$ cm, which is still much smaller than observational value.

Our simulations underestimate the radii of dust grains around V1309 Sco.
The main reason may come from the different environments
between the stellar wind of AGB stars and the binary-merger ejecta. Larger grains require the process of
dust formation and growth in a more stable environment\citep{Nicholls2013}.

\begin{figure}
\includegraphics[totalheight=3.3in,width=3.0in,angle=-90]{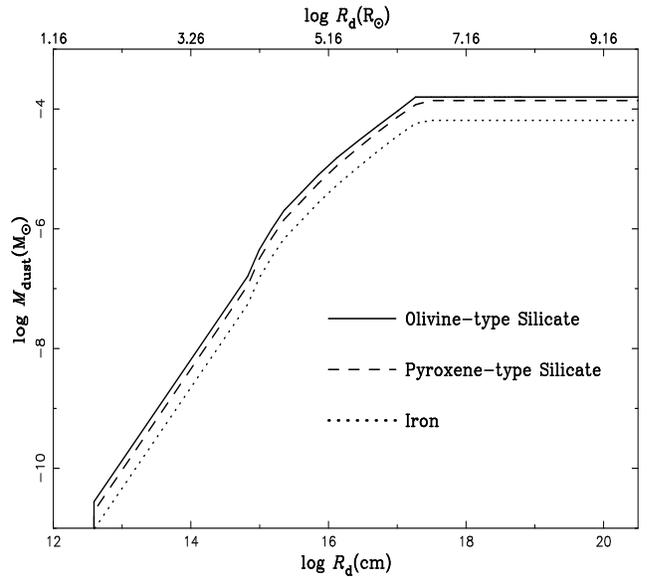}
\caption{The amounts of dust produced in the binary-merger ejecta as a function
of the distance to V1309 Sco.}\label{fig:rqual}
\end{figure}

Fig.\ref{fig:rqual} gives the amounts of dust produced in the binary-merger ejecta as a function
of the distance to V1309 Sco. Based on our models, majority of dust grains form in
the region of $\sim 10^{13}$ --- $10^{17}$
cm away from V1309 Sco. \cite{Nicholls2013} found a significant amount of dust around
V1309 Sco between 18 and 23 months after outburst. \cite{Mason2010} discovered that
V1309 Sco spectra were characterized by low velocities (The
emission lines \emph{\textbf{FWHM}} and their extended wings never exceeded
150 km s$^{-1}$ and 1000 km s$^{-1}$, respectively).
Similarly, simulating the merging process of the progenitor binary of V1309 Sco,
\cite{Ivanova2013} showed that some of the ejecta only barely faster than the local escape velocity (It is $\sim 420$ km s$^{-1}$)
and some will get significantly more specific kinetic energy. If we assume that the ejecta velocity
is $\sim $100---1000 km s$^{-1}$, dust grains discovered by \cite{Nicholls2013} should be located in a zone of
$\leq 10^{14}$ --- $10^{16}$ cm away from V1309 Sco. This is agreement with our results.

\section{Conclusions}
The discovery of a significant amount of dust around V1309 Sco directly supports
the point of view in \cite{Lu2013}, that is, CE ejecta in a
close binary system provides an ideal environment for dust
formation and growth. Using their toy model, we find that the ejecta of
progenitor-binary merger can efficiently produce dust and there are $\sim 5.2\times10^{-4} M_\odot$
of dust grains around V1309 Sco.

Based on \cite{Han1995a},  about 20\% of
all binary systems undergo CE evolutions, and about 40\% of these CE evolutions result in binary mergers.
However, as Fig. \ref{fig:tede} shows, only tiny part of whole envelope is ejected. Compared with CE ejecta,
the contribution of dust produced by binary-merger ejecta to the overall dust
production in the ISM is negligible.

\section*{Acknowledgments}
GL thanks H.-P. Gail for offering \emph{AGBDUST} code. This work was
supported by the National Natural Science Foundation of China under
Nos. 11363005, 11063002 and 11163005, Foundation of Ministry of Education under No.
211198 and Foundation of Huoyingdong under No.
121107.
\bibliographystyle{apj}
\bibliography{lglapj}


\label{lastpage}

\end{document}